\documentclass[aps,pre,showpacs,preprintnumbers,amsmath,amssymb,twocolumn,superscriptaddress]{revtex4-1}
\usepackage{graphicx}
\usepackage{bm}
\usepackage{amssymb}
\usepackage{array}
\usepackage{longtable}
\usepackage{dcolumn}
\usepackage{hyperref}
\usepackage{float}
\usepackage{natbib}
\usepackage{color}
\usepackage{ulem}

\newcommand{\ve}[1][K]{\mathbf{#1}}

\begin{document}

\title{Contact Kinetics in Fractal Macromolecules}

\author{Maxim Dolgushev}
\affiliation{Physikalisches Institut, Universit\"at Freiburg, Hermann-Herder-Str. 3, 79104 Freiburg, Germany}
\author{Thomas Gu\'erin}
\affiliation{Universit\'e de  Bordeaux and CNRS, Laboratoire Ondes et Mati\`ere d'Aquitaine (LOMA), UMR 5798, 33400 Talence, France}
\author{Alexander Blumen}
\affiliation{Physikalisches Institut, Universit\"at Freiburg, Hermann-Herder-Str. 3, 79104 Freiburg, Germany}
\author{Olivier B\'enichou}
\author{Rapha\"el Voituriez}
\affiliation{Laboratoire de Physique Th\'eorique de la Mati\`ere Condens\'ee, CNRS/UPMC, 
 4 Place Jussieu, 75005 Paris, France}

\begin{abstract}
We consider the  kinetics of first contact between two monomers of the same macromolecule. Relying on a fractal description of the macromolecule, we develop an analytical method to compute the Mean First Contact Time (MFCT) for various molecular sizes. In our theoretical description, the non-Markovian feature of monomer motion, arising from the interactions with the other monomers, is captured by accounting for the  non-equilibrium conformations of the macromolecule at the very instant of first contact. This analysis reveals  a simple scaling relation for the MFCT between two monomers, which involves only their equilibrium  distance  and  the spectral dimension of the macromolecule, independently of its microscopic details. Our theoretical predictions are in excellent agreement with  numerical stochastic simulations. 
\end{abstract}


\date{\today}
\maketitle

\textit{Introduction.} Intramolecular reactions are ubiquitous in nature. Examples are provided by the formation of RNA hairpins \cite{Bonnet1998,*Wang2004} or DNA loops \cite{Wallace2001,*Allemand2006}, the folding of polypeptides \cite{Lapidus2000,*Moeglich2006}, as well as the appearance of cycles in synthetic polymers \cite{Gooden1998,*Zheng2011,*Burgath2000}. It is generally known that the structure of a macromolecule has a strong influence on the dynamics of its monomers \cite{DoiEdwardsBook,Gurtovenko2005}, and that this complex intramolecular dynamics implies nontrivial reaction kinetics \cite{DEGENNES1982} in the diffusion limited regime. Until now, most of the theoretical work  devoted to the reaction times in macromolecules has been limited mainly to linear chains  \cite{Wilemski1974a,*Wilemski1974b,Szabo1980,DEGENNES1982,Friedman1993,Sokolov2003,Guerin2012,Amitai2012,Afra2013,Guerin2013a,*Guerin2013b,*Guerin2014,*Guerin2015,Shin2015a,*Shin2015b}. However, there are numerous examples of macromolecules which differ from  linear polymer chains \cite{Voit,Burchard1999}; their dynamic and static properties often suggest a fractal character \cite{Burchard1999,Sokolov2002,Gurtovenko2005}. In particular, the dynamics of fractal macromolecules is characterized by dynamical exponents that are
different from those of linear chains \cite{Sokolov2002,Blumen2004,Gurtovenko2005,Reuveni2008,*Reuveni2010,*Reuveni2012}, leading presumably to distinct first contact kinetics that cannot be deduced from existing works on linear chains. It is important to note the significance of fractals: They provide typical models, e.g., for hyperbranched polymers \cite{Gurtovenko2005}, proteins \cite{Burioni2004,*Banerji2011,Reuveni2008,*Reuveni2010,*Reuveni2012}, sol-gel branched clusters \cite{Zilman1998}, and colloidal aggregates \cite{Sorensen2001}. 

\begin{figure}
\includegraphics[width=0.9\linewidth]{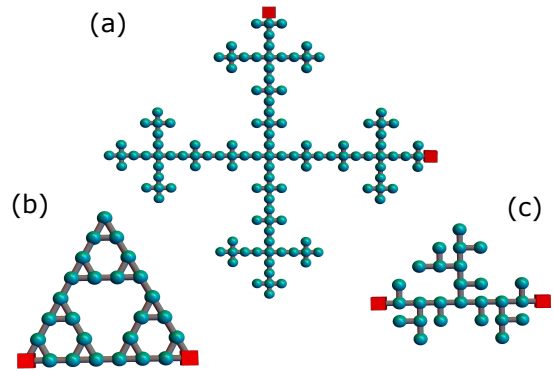}
\caption[kurzform]{(color online) Structure of fractal macromolecules investigated in this Letter: (a) Vicsek fractal, here of functionality (i.e. number of nearest-neighbors of the branching sites) $f=4$ (VF4), (b) dual Sierpi\'{n}ski gasket (DSG), (c) $T-$fractal (TF). The reactive monomers for which we compute the MFCT are represented by red squares. These extended conformations show only the topology of the structures. 
}\label{CartoonFractal}
\end{figure}

The aim of this Letter is to propose a theoretical description of the mean first contact time (MFCT) between two monomers  in a fractal macromolecule, described as a network of beads connected by springs. The cornerstone feature of such  fractal models is the anomalous vibrational dynamics of the network \cite{Gurtovenko2005,Blumen2004,Reuveni2008,*Reuveni2010,*Reuveni2012}. It originates from the non-Debye density of states which is characterized through the so-called spectral dimension $d_s$ \cite{Nakayama1994} (also known as "fracton" dimension \citep{Alexander1982}). In this Letter, we go beyond existing studies that focused on the specific case of linear chains only \cite{Wilemski1974a,*Wilemski1974b,Szabo1980,DEGENNES1982,Friedman1993,Sokolov2003,Guerin2012,Amitai2012,Afra2013,Guerin2013a,*Guerin2013b,*Guerin2014,*Guerin2015,Shin2015a,*Shin2015b}, and show on general grounds that $d_s$ (rather than the microscopic properties of the macromolecule) is the key parameter that controls intramolecular reaction kinetics. Indeed, the spectral dimension $d_s$ will be shown to leave its fingerprint in the scaling behavior of MFCTs with the equilibrium distance between monomers, as  will be confirmed by explicit computations on examples of fractal structures (see Fig.~\ref{CartoonFractal}).

It is important to stress that in the case of macromolecules, the interactions between monomers lead to an effective non-Markovian motion, which is the hallmark of monomer dynamics \cite{Panja2010,*Gupta2013}. In this Letter, we take into account such non-Markovian features and describe the contact kinetics for fractal structures. We show explicitly that the non-Markovian effects increase with the complexity and the degree of branching of the macromolecules.

\textit{Model.} The macromolecular structures are represented by $N$ beads located at positions $\mathbf{r}_i(t)$ in a 3-dimensional space and connected by springs of stiffness $K$. The free-draining dynamics of the structure is given by the Langevin equations~\cite{DoiEdwardsBook,Gurtovenko2005}:
 \begin{equation}
\label{langevin}
 	\zeta \frac{\partial}{\partial t}\ve[r]_{i}(t)+K \sum_{j=1}^{N}A_{ij}\ve[r]_{j}(t)=\ve[F]_{i}(t),
\end{equation}
where $\mathbf{A}=(A_{kj})$ is the connectivity (Laplacian) matrix that describes the topology of the structure. The off-diagonal elements $A_{ij}$ are equal to $-1$ if beads $i$ and $j$ are connected, and $0$ otherwise. For each bead $i$, the diagonal element $A_{ii}$ is equal to the number of bonds emanating from it. Also Eq. (\ref{langevin}) includes friction forces $-\zeta\partial_t\ve[r]_i$ and stochastic forces $\mathbf{F}_i(t)$ obeying white noise statistics with amplitude $\langle F_{i\alpha}(t)F_{j\beta}(t')\rangle=2k_{B}T\zeta\delta(t-t')\delta_{ij}\delta_{\alpha\beta}$, where $k_BT$ represents the thermal energy and $\alpha,\beta$ are spatial coordinates $x,y,z$. It is natural to introduce the monomeric relaxation time $\tau_0=\zeta/K$, and the characteristic microscopic length $l=(3 k_BT/K)^{1/2}$ (in the case of structures without loops $l^2$ is the mean-squared bond length).

Here we consider the contact kinetics between two given monomers denoted "reactive monomers", whose indexes are called $q_1$ and $q_2$ (see  Fig. \ref{CartoonFractal}). 
We introduce the vector $\ve[R](t)$ that joins them:
\begin{align}
	\ve[R](t)\equiv \ve[r]_{q_1}(t)-\ve[r]_{q_2}(t)\equiv \sum_{i=1}^N h_i \ve[r]_i(t), \label{defXandh}
\end{align}
where $h$ is a $N-$dimensional vector defined by this equation, which has only 2 non-zero elements in positions $q_1$ and $q_2$. It is convenient to decompose the Gaussian vector $\ve[R](t)$ as a sum of independent modes,  
\begin{align}
	\ve[R](t)=\sum_{\lambda} b_{\lambda} \ve[a]_{\lambda}(t), \label{94842}
\end{align}
where $\lambda$ represents all the distinct nonvanishing eigenvalues of $\ve[A]$, $b_{\lambda}^2$ is the norm of the orthogonal projection of the vector $h$ [defined in Eq. (\ref{defXandh})] on the eigensubspace associated to $\lambda$ \footnote{See Supplemental Material at [URL will be inserted by
publisher] for the identification of $b_{\lambda}^2$ and the iterative calculation of correlation
functions, as well as for the details of the MFCT theories, which includes Refs.\cite{Meyer2000,Guerin2013a,Sommer1995,Reuveni2012pre,Blumen2004,Fuerstenberg2013,Rammal,Cosenza}}. The $\ve[a]_{\lambda}(t)$ evolve independently of each other with the correlation function:
\begin{align}
\langle a_{\lambda,\alpha}(t)a_{\lambda',\beta}(t')\rangle=l^2\delta_{\alpha,\beta}\delta_{\lambda,\lambda'}e^{-\lambda\vert t-t'\vert/\tau_0}/(3\lambda) \label{904301}.
\end{align}
In this picture of independent modes, the normalized temporal autocorrelation function of the Cartesian components $(R_x(t), R_y(t), R_z(t))$ of vector $\ve[R](t)$ follows from Eqs.~(\ref{94842}-\ref{904301}):
\begin{align}
	\phi(t)\equiv \frac{\langle R_{\alpha}(t) R_{\alpha}(0)\rangle}{\langle R_{\alpha}(0)^2\rangle } =\left(\sum_{\lambda} \frac{b_{\lambda}^2}{\lambda} e^{-\lambda t/\tau_0}\right)\Big/\sum_{\lambda} \frac{b_{\lambda}^2}{\lambda} .
\end{align}
Since $\ve[R](t)$ is a Gaussian stochastic process, it is entirely characterized by the function $\phi(t)$. We also introduce $n=\langle [\ve[R](t)/l]^2\rangle=\sum_{\lambda}b_{\lambda}^2/\lambda$, which for macromolecules devoid of loops can be shown to be  simply the number of bonds connecting the reactive monomers. 

\begin{figure}
\includegraphics[width=0.95\linewidth,trim=0 0 0 0]{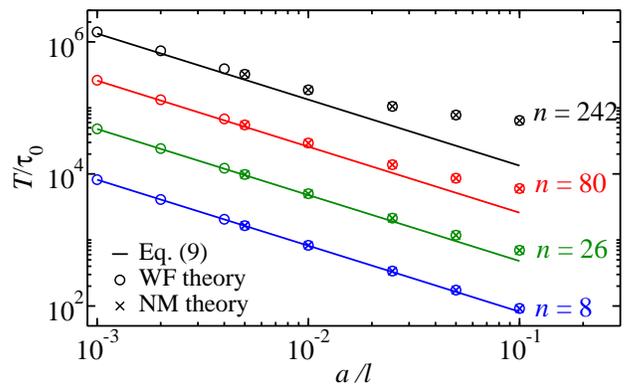}
\caption[kurzform]{(color online) $T_{\text{NM}}$ (Eq.~(\ref{ExpressionTNonMarkovianCentre})) and $T_{\text{WF}}$ (Eq.~(\ref{ExpressionTMarkovianCentre})) for Vicsek fractals of functionality $f=4$ as a function of the capture radius $a$. The lines represent the results of Eq.(\ref{eq_tu}).}\label{T_asmall}
\end{figure}

\textit{Theories of first contact times.} We now sketch briefly the method we use for the calculation of the MFCT, defined as the average time needed for the reactive monomers to be separated by a distance smaller than $a$ (called the capture radius), starting from an initial equilibrium configuration in which the reactive monomers are not in contact. We introduce the joint probability density  $f(\{\ve[a]\}, t)$ that contact is made for the first time at time $t$ and that, at this first passage event, the macromolecule has a configuration described by the set of modes $\{\ve[a]\}=(\ve[a]_1,\ve[a]_2...)$. Let us partition the trajectories that lead to a configuration $\{\ve[a]\}$ (in which the contact condition is satisfied) into two steps, the first step consisting in reaching the target for the first time at $t'$, and the second step consisting in reaching the final configuration $\{\ve[a]\}$ in a time $t-t'$. The mathematical formulation that corresponds to this decomposition of events is 
\begin{align}
	p(\{\ve[a]\},t)=
&\int_0^t dt' \int d\{\ve[a]'\} f(\{\ve[a]'\},t') p(\{\ve[a]\},t-t'\vert \{\ve[a]'\}),\label{EquationRenewal}
\end{align}
where $p(\{\ve[a]\},t\vert \{\ve[a']\})$ is the probability of $\{\ve[a]\}$ at $t$ starting from $ \{\ve[a']\}$ at $t=0$ while $p(\{\ve[a]\},t)$ is the probability of $\{\ve[a]\}$ at $t$ starting from the initial conditions. Although Eq. (\ref{EquationRenewal}) is exact, it is in general very difficult to solve. A classical approximation, introduced by Wilemski and Fixman (WF) \cite{Wilemski1974a,*Wilemski1974b}, assumes that $f(\{\ve[a]\},t)$ is proportional to the equilibrium distribution of configurations that satisfy the constraint that a contact is formed.  Introducing this approximation into Eq. (\ref{EquationRenewal}), integrating over all configurations and taking the long time limit lead to the estimate $T_{\mathrm{WF}}$ of the mean first contact time \cite{Guerin2013a}
\begin{align}
	T_{\text{WF}}=\int_0^{\infty} dt\left\{\frac{ e^{-a^2\phi(t)^2/[2\psi(t)]}}{[1-\phi(t)^2]^{3/2}}- \frac{Z(a/\psi(t)^{1/2})}{Z(a\sqrt{3}/l\sqrt{n})}
\right\},\label{ExpressionTMarkovianCentre}
\end{align}
where $\psi(t)=n l^2[1-\phi(t)^2]/3$ is the mean-square displacement of $R_x(t)$ and $Z(y)=\int_y^{\infty}dx \ x^2e^{-x^2/2}$.
While this approximation takes into account some aspects of the complex dynamics of the macromolecule through the correlation function $\phi$, it neglects non-Markovian (NM) effects which can be quantitatively important. These NM effects can be described by considering the distribution of configuration at the first contact event $\pi(\{\ve[a]\})=\int_0^{\infty}dt f(\{\ve[a]\}, t)$. The analytic expression for $\pi(\{\ve[a]\})$ is unknown; following the case of linear polymers~\cite{Guerin2012}, we assume that it is well approximated by a Gaussian distribution, which is therefore characterized by its first and second moments. Its first moments, denoted $m_{\lambda}$, are the average mode amplitudes $\ve[a]_{\lambda}$ at the first contact instant $t^*$ (in the direction of the vector 
$\ve[R](t^*)$), while the second moments can be approximated by their equilibrium values. Then, a precise estimate of the MFCT is obtained by integrating  (\ref{EquationRenewal}) over all contact configurations, leading to
\begin{align}
	&T_{\text{NM}}=\int_0^{\infty} dt\left\{\frac{ e^{-R_{\pi}(t)^2/[2\psi(t)]}}{[1-\phi(t)^2]^{3/2}}-  
\frac{Z(a/\psi(t)^{1/2})}{Z(a\sqrt{3}/l\sqrt{n})}
\right\}.\label{ExpressionTNonMarkovianCentre}
\end{align}
The difference between this expression and Eq. (\ref{ExpressionTMarkovianCentre}) lies in the presence of the reactive trajectory $R_{\pi}(t)$, defined as the average of $\ve[R]$ at a time $t$ after the first contact $t^*$ in the direction of $\ve[R](t^*)$, and thus reads  $R_{\pi}(t)=\sum_{\lambda} b_{\lambda} m_{\lambda} e^{-\lambda t/\tau_0}$. The involved  $m_{\lambda}$  are obtained from a set of self-consistent equations, see the Supplemental  Material~\cite{Note1} for details.

\textit{Cyclization for small capture radius.} In the limit of $a\rightarrow0$ (while keeping a fixed number of monomers), only the dynamics at small time scales matters, where $\psi(t)\simeq 4 D_0 t$ with the local diffusion constant $D_0=k_BT/\zeta$. Also, at short times, we can write $R_{\pi}(t)\simeq R_{\pi}(0)=a$. Introducing these approximations into Eqs. (\ref{ExpressionTMarkovianCentre},\ref{ExpressionTNonMarkovianCentre}), we deduce that both the WF and the NM theories predict a MFCT which reads
\begin{equation}
	T\simeq \tau_0\  \pi^{1/2}n^{3/2}l/(2\sqrt{6}\ a)  \hspace{0.5cm}(\text{for }a\rightarrow0).\label{eq_tu}
\end{equation}
Thus, in this regime the MFCT is independent of the particular polymeric structure, and has the same expression as in the case of linear chains \citep{Guerin2012,Guerin2013a,Guerin2014}. The convergence of both Eqs.~(\ref{ExpressionTMarkovianCentre},\ref{ExpressionTNonMarkovianCentre}) to the scaling form (\ref{eq_tu}) for small capture radius is demonstrated in Fig.~\ref{T_asmall}.

\textit{Scaling for large $n$. } When the size of the structure grows, Eq. (\ref{eq_tu}) is not valid anymore and the MFCT reflects then the complex monomer dynamics, which becomes subdiffusive, $\langle [\ve[r]_q(t)-\ve[r]_q(0)]^2\rangle\sim t^{\gamma}$, with $\gamma$ a subdiffusive exponent related to the spectral dimension $d_s$ of the structure, $\gamma=1-d_s/2$  for $d_s<2$ \cite{Alexander1982,Nakayama1994,Blumen2004,Sommer1995}. We first present a simple qualitative argument  to derive the behavior of the MFCT as a function of $n$, which we explicitly check for the WF theory and validate next for the NM theory by a numerical analysis of the analytical results of Eq.~(\ref{ExpressionTNonMarkovianCentre}).

For a Markovian subdiffusive walker whose subdiffusive exponent is $\gamma$, the time needed to find a point-like target in a confining volume of size $L$ starting at a random position is proportional to $L^{2/\gamma}$ \citep{Condamin2007}. Given that in our case the typical length scale is $L\sim\sqrt{n}$, the scaling argument 
\begin{align}
T\sim \tau_0 \ n^{1/\gamma} \label{ScalingMFCT}
\end{align}
follows. We do not expect a strong dependence on the capture radius since the dynamics of a monomer is compact (or recurrent) \cite{DEGENNES1982,Condamin2007}. In fact, the scaling (\ref{ScalingMFCT}) can be derived in the WF approximation by noting that for extremal monomers (i.e. monomers whose relative distance is maximal), the time $\tau_0 \ n^{1/\gamma}$ is also of the order of the maximal relaxation time $\tau_N$ \cite{Alexander1982,Reuveni2012pre}, such that the correlation function scales as $\phi(t)=\tau_N\Phi(t/\tau_N)$. Once this equality is reported into Eq.~(\ref{ExpressionTMarkovianCentre}) one obtains that $T\sim \tau_N$, which is the expected behavior (\ref{ScalingMFCT}), see  the Supplemental  Material~\cite{Note1} for details. Strikingly, the MFCT is found to depend on the polymer structure through $d_s$ only (and not on its microscopic details). In particular, this shows that the MFCT of highly branched structures ($\gamma\rightarrow0$) differs significantly from that of linear chains ($\gamma=1/2$). 

\begin{figure}
\includegraphics[width=0.95\linewidth,trim=0 0 0 0]{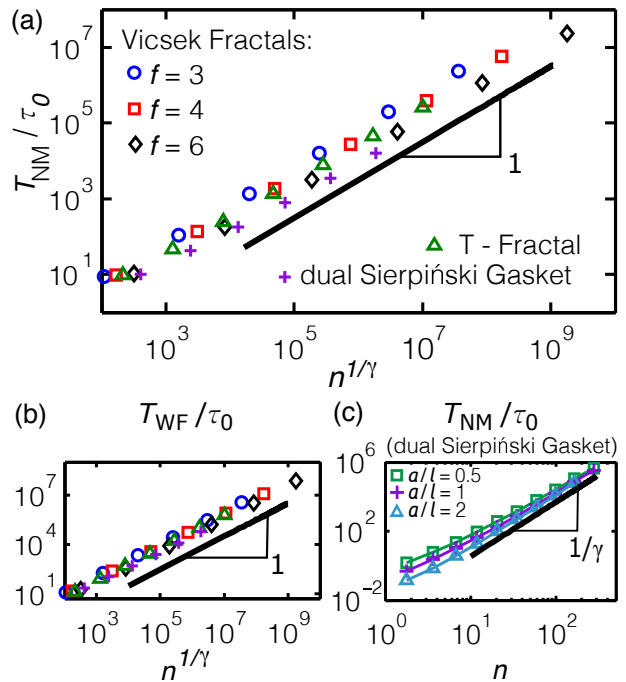}
\caption[kurzform]{(color online) (a) NM and (b) WF  MFCT for different fractal structures as a function of $n^{1/\gamma}$. The parameter $\gamma$ is obtained from the known values \cite{Blumen2004,Cosenza,Agliari2008} of the spectral dimension $d_s$: for Vicsek fractals of functionality $f$, $\gamma=\ln(3)/\ln(3f+3)$, for dual Sierpi\'{n}ski gasket $\gamma=\ln(5/3)/\ln(5)$, and for T-fractal $\gamma=\ln(2)/\ln(6)$. The capture radius is $a=l$. (c) NM MFCT for the dual Sierpi\'{n}ski gasket for different values of $a$.} \label{T_a1}
\end{figure}

Now, checking the scaling (\ref{ScalingMFCT}) requires the actual computation of the correlation function $\phi(t)$, which itself involves the diagonalization of $\ve[A]$. However, a naive diagonalization of this matrix (as usually done for  linear chains, where the analytic diagonalization is possible) does not allow to deal with structures large enough.
This difficulty can be overcome by exploiting the highly symmetric nature of the fractal macromolecules we are considering. First, we remark that the actual number of variables needed to be taken into account is the number of \textit{distinct} eigenvalues, which is much lower than the total number of beads $N$. Then, one can set up a decimation procedure inspired by that used to find iterative formulas for the eigenvalues \cite{Blumen2004,Meyer2012}. Adapting this decimation approach 
to the iterative computation of the coefficients $b_{\lambda}^2$ amounts to solving a linear algebra problem, as described in the Supplemental  Material~\cite{Note1}. In practice, by using the WF and NM formalisms we were able to calculate the MFCT for macromolecules containing as many as $800,000$ beads, which would not have been possible through direct diagonalization.  

Based on the iterative computation of $b_{\lambda}^2$, we test the scaling behavior (Fig.~\ref{T_a1}) for different fractal structures (represented in Fig.~\ref{CartoonFractal}) and therefore for several values of the subdiffusive exponent $\gamma$ (or $d_s$). As can be seen on Fig.~\ref{T_a1}, for all these structures, the scaling $T\sim n^{1/\gamma}$ is in good agreement with the predictions of both the NM theory and the WF theory. This confirms that the spectral dimension $d_s$ of the structure plays a fundamental role for MFCT. In particular, the functionality $f$, which determines $\gamma$, plays a crucial role and yields scaling behaviors that can differ significantly from those of linear chains. However, the presence of many loops in the dual Sierpi\'{n}ski gasket does not modify the scaling behavior of the MFCT: In Fig. \ref{T_a1}(c) we show the results of the corresponding MFCT $T_{\text{NM}}$ for 3 different values of $a$. We observe that all curves scale in the same way for large $n$, independently of the capture radius, fact consistent with Eq.~(\ref{ScalingMFCT}). 
 
\textit{Comparison with numerical simulations. }
We checked the validity of our theoretical predictions by performing Brownian dynamics simulations of Eq. (\ref{langevin}), using the algorithm of Ref.~\cite{Pastor1996} with fixed time steps. The results are presented in Table~\ref{table_sim} for different fractal structures and capture radia $a$. We observe that the WF theory systematically overestimates the MFCT, whereas the NM theory describes the simulation data almost quantitatively, thereby validating the accuracy of our analysis. 

\begin{table}
\begin{ruledtabular} 
\begin{tabular}{cccccc}
$a/l$ & structure & $1/\gamma$ & $T_{\mathrm{WF}}$ & $T_{\mathrm{NM}}$ & $T_{\mathrm{simu}}$ \\\hline
$1$ & DSG & $3.15$ & $3.85$ & $2.38$ & $2.41\pm0.02$ \\
& TF & $2.58$ & $13.31$ & $9.48$ & $9.53\pm0.07$ \\
& VF$3$ & $2.26$ & $166.8$ & $111.5$ & $113.6\pm0.9$ \\
& VF$4$ & $2.46$ & $224.9$ & $135.1$ & $137.4\pm1.4$ \\
$2.5$ & VF$3$ & $2.26$ & $87.81$ & $37.52$ & $37.47\pm0.35$ \\
& VF$4$ & $2.46$ & $128.3$ & $47.1$ & $45.3\pm0.5$ \\
\end{tabular} 
\end{ruledtabular} 
 \caption{MFCT for the structures of Fig.~\ref{CartoonFractal} computed with the WF ($T_{\mathrm{WF}}$) and NM ($T_{\mathrm{NM}})$ theories [with Eqs. (\ref{ExpressionTMarkovianCentre},\ref{ExpressionTNonMarkovianCentre})], compared to the results $T_{\mathrm{simu}}$ of the stochastic simulations of (\ref{langevin}). $1/\gamma$ indicates numerical values of the scaling exponent in (\ref{ScalingMFCT}) (in the case of linear chains it is equal to $2$ \cite{Wilemski1974a,*Wilemski1974b,Szabo1980,Guerin2012}). All structures are of generation $g=3$. The times are units of $\tau_0$. }\label{table_sim} 
\end{table} 

\begin{figure}
\includegraphics[width=0.9\linewidth]{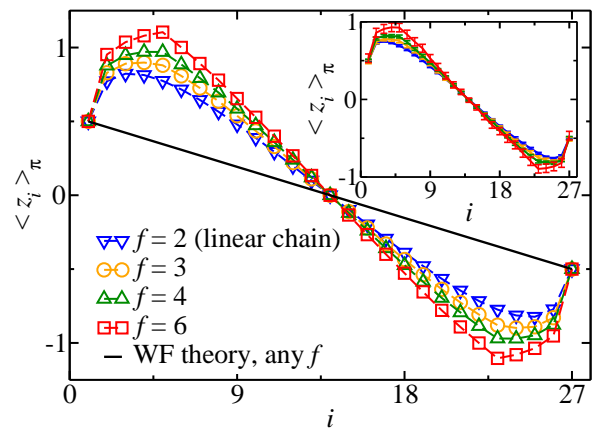}
\caption[kurzform]{(color online) Average spatial position $\langle z_i\rangle_{\pi}$ of the monomers (numbered through $i=1\dots n$) connecting reactants at the instant cyclization in the direction of the reaction, for Vicsek fractals of different functionalities $f$ with generation $g=3$, as predicted by the NM (symbols) and WF theories (solid line). Inset: same quantity determined from simulations, with the same color code.}\label{positions}
\end{figure}

\textit{Average conformations at first contact. }
Inspecting Fig. \ref{T_a1} reveals that the WF theory overestimates the MFCT by a numerical factor which grows when $\gamma$ decreases. This fact is also clearly seen in Table \ref{table_sim}, and confirmed by the numerical simulations. This means that, with decreasing subdiffusive exponent $\gamma$, the macromolecular conformations at the instant of first contact differ more and more from equilibrium ones. To illustrate this fact, we present on Fig.~\ref{positions} the average spatial positions of the monomers in Vicsek fractals at the instant of first contact $\langle z_i\rangle_{\pi}\equiv\langle \ve[r]_i(t^*)\cdot\hat{\ve[u]}(t^*)\rangle$, where $t^*$ represents the first contact instant, $\hat{\ve[u]}(t^*)=\ve[R](t^*)/\vert\ve[R](t^*)\vert$ gives the direction between the reactive monomers at $t^*$ and $i$ are the indices of the beads that lie between the reactive monomers (on the chain). In the equilibrium WF theory, all monomers between the reactants (on the chain) are on average also between the reactants in space at the instant of first contact, irrespectively of the functionality of the macromolecule. 
The NM theory, in turn, predicts that those monomers which are close to the reactants are in average outside the capture radius at $t^*$ (Fig.~\ref{positions}, symbols). Moreover, this effect is more pronounced for Vicsek fractals of higher functionalities, meaning that NM effects increase with the degree of hyperbranching of the macromolecule. As shown in the inset of Fig. \ref{positions}, the simulations confirm these conclusions, although the value of $\langle z_i\rangle_{\pi}$ is slightly overestimated in the NM theory. 

\textit{Conclusion.} 
Summarizing, we have studied the kinetics of first contact between two monomers belonging to the same fractal macromolecule. We identified two regimes: (i) for very small capture radius, the MFCT becomes independent of the macromolecular structure and originates essentially from the microscopic diffusive motion of the monomers, whereas (ii) for larger capture radius, the MFCT scales as a power-law with the mean-square distance $n$ between the monomers, with an exponent related to the spectral dimension $d_s$ and independent of microscopic details. We confirmed this scaling law for a wide variety of structures (with and without loops). The non-Markovian effects are included by calculating the average equilibrium conformation of the whole macromolecule at the instant of first contact, and are found to be more important when the degree of hyperbranching of the structures increases. Finally, it would be interesting to explore how the  MFCT varies with the location of the reactants in the structure or to generalize the theory to include the effect of hydrodynamic interactions, which typically would lead to new scalings \cite{Reuveni2012pre}.

\begin{acknowledgments} \textit{Acknowledgments.} M.D. and A.B. acknowledge the support of the DAAD through the PROCOPE program (Project No. 55853833), of the DFG through Grant No. Bl 142/11-1 and through IRTG “Soft Matter Science” (GRK 1642/1). T.G., O.B., and R.V. acknowledge the support of Campus France (project No. 28252XE) and of the European Research Council starting Grant FPTOpt-277998.
\end{acknowledgments}

\vspace{1cm}
{\bf Supplementary information}

\appendix
\section{Determination of the autocorrelation functions}

Here we present general expressions for the determination of the temporal autocorrelation function of the vector $\mathbf{R}(t)$ connecting the reactive monomers $q_1$ and $q_2$, defined as
\begin{align}
	\ve[R](t)\equiv \ve[r]_{q_1}(t)-\ve[r]_{q_2}(t)\equiv \sum_{i=1}^N h_i \ve[r]_i(t) \label{defXandh}.
\end{align} 
where $\mathbf{h}=(h_i)$ is the $N$-components vector 
\begin{align}
h_i=\delta_{i,q_1}-\delta_{i,q_2}.\label{Def_Vector_h}
\end{align} 
We recall that the positions $\{\ve[r]_i(t)\}$ of the $N$ beads in the macromolecule obey the Langevin equation [Eq.(1) in the main text]:
\begin{equation}\label{langevin}
 	\zeta \frac{\partial}{\partial t}\ve[r]_{i}(t)+K \sum_{j=1}^{N}A_{ij}\ve[r]_{j}(t)=\ve[F]_{i}(t),
\end{equation}
where $\mathbf{A}$ is the connectivity matrix that describes the topology of the structure. The off-diagonal elements $A_{ij}$ are equal to $-1$ if beads $i$ and $j$ are connected, and to $0$ otherwise. For each bead $i$, the diagonal element $A_{ii}$ is equal to the number of bonds emanating from it. Finally, in Eq. (\ref{langevin}), the stochastic forces $\mathbf{F}_i(t)$ follow the statistics
\begin{align}
\langle F_{i\alpha}(t)F_{j\beta}(t')\rangle=2\ k_{B}T\ \zeta\ \delta(t-t')\delta_{ij}\delta_{\alpha\beta},
\end{align}
where $k_BT$ represents the thermal energy and $\alpha,\beta$ are the spatial coordinates $x,y,z$. 
It is natural to introduce the microscopic time scale $\tau_0=\zeta/K$. 

We introduce the  eigenvectors of the matrix $\mathbf{A}$ and denote by $u_i^{(\lambda,q)}$ the $i$th coordinate of the $q$th eigenvector associated with the eigenvalue $\lambda\neq0$. If $\lambda$ has degeneracy $p_{\lambda}$, then $q$ is bounded by $1\leq q\leq p_{\lambda}$. Since the matrix $\ve[A]$ is symmetric and real, we can assume that these eigenvectors are orthogonal and normalized. It is then standard to decompose the motion of each bead in the structure in a sum of eigenmode amplitudes, 
\begin{align}
\ve[r]_i(t)=\sum_{\lambda}\sum_{q=1}^{p_{\lambda}}u_i^{\lambda,q}\ \tilde{\ve[a]}_{\lambda,q}(t). \label{0592}
\end{align}
Inserting (\ref{0592}) into (\ref{langevin}) and taking into account the fact that the vectors $u_i^{\lambda,q}$ are orthogonal, we obtain that all the eigenmode amplitudes evolve independently of each other, with the correlation function
\begin{align}
\langle \tilde{a}_{\lambda,q,\alpha}(t)\tilde{a}_{\lambda',q',\beta}(t')\rangle=\frac{ k_BT}{ \lambda K} \delta_{\alpha,\beta}\delta_{q,q'}\delta_{\lambda,\lambda'}\ e^{-\lambda\vert t-t'\vert/\tau_0}. \label{CorrAlambdaq} 
\end{align}
Let us introduce now, for each eigenvalue $\lambda$, the positive coefficient $b_{\lambda}$ and the stochastic variable $\ve[a]_{\lambda}(t)$ defined as 
\begin{align}
&b_{\lambda}^2=\sum_{q=1}^{p_{\lambda}}\left(\sum_{i=1}^{N} h_i u^{(\lambda,q)}_i\right)^2 \label{DefBLambda},\\
&\ve[a]_{\lambda}(t)=\frac{1}{b_{\lambda}}\sum_{q=1}^{p_{\lambda}}\sum_{i=1}^{N} h_i u^{(\lambda,q)}_i \tilde{\ve[a]}_{\lambda,q}(t).\label{Def_alambda}
\end{align}
Comparing (\ref{Def_alambda}) and  (\ref{CorrAlambdaq}), it is clear that the autocorrelation function of the stochastic variables $\ve[a]_{\lambda}$ is
\begin{align}
	\langle a_{\lambda,\alpha}(t)a_{\lambda',\beta}(t')\rangle=\frac{k_BT}{K\lambda} \delta_{\alpha,\beta}\delta_{\lambda,\lambda'}e^{-\lambda\vert t-t'\vert/\tau_0},
\end{align}
which is exactly Eq. (4) of the main text. 
Furthermore, it follows from (\ref{defXandh},\ref{0592},\ref{DefBLambda}) that
\begin{align}
	\ve[R](t)=\sum_{\lambda}\sum_{q=1}^{p_{\lambda}}\sum_{i=1}^{N} h_i u^{(\lambda,q)}_i \tilde{\ve[a]}_{\lambda,q}(t) = \sum_{\lambda}b_{\lambda}\ve[a]_{\lambda}(t)
\end{align} 
which is Eq. (3) of the main text. Finally, we note that (\ref{DefBLambda}) can be written as
\begin{equation}\label{b_lambda3}
b_{\lambda}^2=\sum_{q=1}^{p_{\lambda}} (\ve[h]| \ve[u]^{(\lambda,q)})^2=(\ve[h]|\hat{P}_{\lambda}|\ve[h]),
\end{equation}
where we have considered $\ve[h]$ and $\ve[u]^{(\lambda,q)}$ as $N-$components vectors, and where the projection operator $\hat{P}_{\lambda}$ is
\begin{equation}\label{def_P}
\hat{P}_{\lambda}=\sum_{q=1}^{p_{\lambda}} |\ve[u]^{(\lambda,q)})(\ve[u]^{(\lambda,q)}|.
\end{equation}
Thus, $b_{\lambda}^2$ is the norm of the orthogonal projection of the $N-$component vector $\mathbf{h}$ on the eigensubspace associated to $\lambda$, as claimed in the main text. 

For the practical calculation of $b_{\lambda}^2$, it is useful to consider a basis of vectors $\{|\ve[w]^{(\lambda,q')})\}$ which span the orthogonal complimentary eigensubspace associated to $\lambda$  (\textit{i.e.} $(\ve[w]^{(\lambda,q')}| \ve[u]^{(\lambda,q)})=0$). The $N-p_{\lambda}$ vectors $\{|\ve[w]^{(\lambda,q')})\}$ are not necessarily orthogonal, but they are linearly independent. 

Introducing the 
$(N-p_{\lambda})\times N$ 
matrix $\mathbf{W}_{\lambda}\equiv(|\ve[w]^{(\lambda,(p_{\lambda}+1))}),\dots,|\ve[w]^{(\lambda,N)}))^T$ and its transpose $\mathbf{W}_{\lambda}^T$ we can construct the complementary projection operator $\hat{Q}_{\lambda}\equiv\hat{I}-\hat{P}_{\lambda}$ (see, \textit{e.g.}, Ref. \cite{Meyer2000}),
\begin{equation}\label{def_Q}
\hat{Q}_{\lambda}=\mathbf{W}_{\lambda}^T(\mathbf{W}_{\lambda}\mathbf{W}_{\lambda}^T)^{-1}\mathbf{W}_{\lambda}.
\end{equation}
Hence, based on Eq.(\ref{def_Q}), Eq.~(\ref{b_lambda3}) transforms to 
\begin{align}\label{b_lambda_final}
b_{\lambda}^2&=(\ve[h]|\hat{I}-\hat{Q}_{\lambda}|\ve[h])\nonumber \\
&=2-(\ve[h]|\mathbf{W}_{\lambda}^T(\mathbf{W}_{\lambda}\mathbf{W}_{\lambda}^T)^{-1}\mathbf{W}_{\lambda}|\ve[h]),
\end{align}
where we have used $(\ve[h]|\ve[h])=2$, see Eq.~(\ref{defXandh}).

\section{Derivation of the equations of the non-Markovian (NM) theory}

Here we briefly describe the main steps that lead to the equations of the NM theory. 

Subtracting the stationary probability distribution $p_s(\{\ve[a]\})$ from Eq.~(6) in the main text, and integrating the result between $t=0$ and $t=\infty$, one obtains  
\begin{align}
&T\ p_s(\{\ve[a]\})=\nonumber\\
&\int_0^{\infty}dt\int d\{\ve[a]'\} \pi(\{\ve[a]'\})[p(\{\ve[a]\},t\vert\{\ve[a]'\})
-p(\{\ve[a]\},t)], \label{054392}
\end{align}
with the unknown quantities are the MFCT ($T$) and the (normalized) probability distribution $\pi(\{\ve[a]\})$ of the configurations at the instant of first contact. The probabilities  $p(\{\ve[a]\},t\vert\{\ve[a]'\})$, $p(\{\ve[a]\},t)$ and $p_s(\{\ve[a]\})$ are Gaussian and admit analytical expressions. Consider now a unit vector $\hat{\ve[u]}$ and let $\pi_{\hat{\ve[u]}}$ be the distribution of modes at the first contact time $t^*$ with the constraint $\ve[R](t^*)=a\hat{\ve[u]}$. In other words, $\pi_{\hat{\ve[u]}}(\{\ve[a]\})\equiv C\ \pi(\ve[a])\delta(\sum_{\lambda}\ve[a]_{\lambda}b_{\lambda} - a\hat{\ve[u]})$, with $C$ a normalization constant. 

In the NM theory, one assumes that $\pi_{\hat{\ve[u]}}$ is a multivariate Gaussian, with mean vector $m_{\lambda}\hat{\ve[u]}$, while the covariance matrix is approximated by that of the equilibrium configurations satisfying the constraint $\ve[R]=a\hat{\ve[u]}$. Then, the $\{m_{\lambda}\}$ are found by solving a set of equations that make the theory self-consistent. These equations are found by  multiplying Eq.~(\ref{054392}) by $\ve[a]_{\lambda}\delta(\sum_{\lambda'} b_{\lambda'} \ve[a]_{\lambda'}-\ve[r]^*)$ for fixed $\lambda$ and $\ve[r]^*$ (satisfying $\vert\ve[r]^*\vert<a$) and integrating over all configurations. Such calculations involve the evaluation of a number of  Gaussian integrals, as well as averages over angular directions $\hat{\ve[u]}$, and are done carefully in the appendix $C$ of Ref. \cite{Guerin2013a}. The main difference between the calculations of Ref. \cite{Guerin2013a} and ours is the fact that, here, the sums are carried out over the distinct values of the eigenvalues $\lambda$ instead of over all the indexes $i$. The resulting equation is 
\begin{align}
&\int_0^{\infty}\frac{dt}{\psi^{5/2} } \Bigg\{ \left[\frac{\mu_{\lambda}^{\pi,0}R_{\pi}}{3}+ 
\frac{b_\lambda\phi (\phi- e^{-\frac{\lambda t}{\tau_0}})}{\lambda } \right]e^{-\frac{R_{\pi}^2}{2\psi}} \nonumber\\
&- \frac{b_\lambda \phi (\phi-e^{-\frac{\lambda t}{\tau_0}}) }{Z_0(a,n l^2/3) \lambda}  \left[Z_0(a,\psi)-\frac{G_0(a,\psi)}{3\psi}
\right] \Bigg\} =0,\label{EquationFirstMomentDim3AveragedSimplified}
\end{align}
where 
\begin{align}
\mu_{\lambda}^{\pi,0}=m_{\lambda}e^{-\frac{\lambda t}{\tau_0}}- \frac{R_{\pi} b_{\lambda}(1-\phi \ e^{-\frac{\lambda t}{\tau_0}})}{\lambda\psi}
\end{align}
and
\begin{align}
Z_0(a,h)=\int_a^{\infty}dx \ x^2 e^{-\frac{x^2}{2h}},\\
G_0(a,h)=\int_a^{\infty}dx \ x^4 e^{-\frac{x^2}{2h}}\label{DefinitionFunctionG}.
\end{align}
By integration by parts, one obtains $Z_0(a,h)-G_0(a,h)/(3h)=-(a^3/3)e^{-a^2/(2h)}$ so that Eq.~(\ref{EquationFirstMomentDim3AveragedSimplified}) can be written in the compact form
\begin{align}
&\int_0^{\infty}dt   \Bigg\{\Bigg[m_{\lambda}e^{-\frac{\lambda t}{\tau_0}}- \frac{R_{\pi} b_{\lambda}(1-\phi \ e^{-\frac{\lambda t}{\tau_0}})}{\lambda\psi} \Bigg]\frac{R_{\pi}e^{-R_{\pi}^2/2\psi}}{3 \ \psi^{5/2}}  \nonumber\\ 
&+    \frac{b_{\lambda}\phi (\phi- e^{-\frac{\lambda t}{\tau_0}})}{\lambda \psi^{5/2}} \Bigg[e^{-\frac{R_{\pi}^2}{2\psi}} -\frac{(a^3/3) e^{-a^2/2\psi}}{ \int_a^{\infty}dr r^2 e^{-\frac{3r^2}{2nl^2}}} \Bigg]\Bigg\}=0.
\label{EquationFirstMomentDim3AveragedSimplified2}
\end{align}
There are as many equations in Eq. (\ref{EquationFirstMomentDim3AveragedSimplified2}) as the number of distinct nonvanishing eigenvalues $\lambda$, so that the values of $m_{\lambda}$ can be determined numerically, enabling the evaluation of the NM contact time with Eq.~(8) of the main text.
Given that for many fractal structures the eigenvalue spectra contain a lot of degenerate eigenvalues, this method allows to treat very large fractal structures, as shown in the main text.

\section{MFCT scaling for large fractal structures}

Let us consider Eq.~(7) of the main text, 
\begin{align}
	T=\int_0^{\infty} dt\left\{\frac{ e^{-a^2\phi(t)^2/[2\psi(t)]}}{[1-\phi(t)^2]^{3/2}}- \frac{Z(a/\psi(t)^{1/2})}{Z(a\sqrt{3}/l\sqrt{n})},
\right\},\label{ExpressionTMarkovianCentre}
\end{align}
where $Z(y)=\int_y^{\infty}dx \ x^2e^{-x^2/2}$, in the case of large $n$. In our case the reactive monomers are at extremities such that $n^{1/2}\sim R_g/ l$ ($R_g  $ is the gyration radius of the whole structure). Thus, the largest time scale is given by the largest relaxation time of the whole structure $\tau_N$. For this it is known~\cite{Sommer1995,Reuveni2012pre} that 
\begin{align}
	\tau_N\sim \tau_0(R_g/l)^{4/(2-d_s)}.
\end{align}
With $\gamma=1-d_s/2$ and with $n^{1/2}\sim R_g/ l$ we get
\begin{align}
	\tau_N\sim \tau_0 n^{1/\gamma}.
\end{align}
 
If the monomers are separated (in physical space) by $n^{1/2}\sim R_g/l$, we can reasonably assume that 
\begin{align}
	\phi(t)=\Phi( t / \tau_N),\label{large_phi}
\end{align}
where $\Phi$ is a dimensionless function independent on $N$, with $\Phi( \tau)\simeq 1 $ at short times, and exponentially decaying for $\tau\gg1$. Now, using that the function $Z(y)$ involved in Eq.~(\ref{ExpressionTMarkovianCentre}) goes for small $y$ as $Z(y)\simeq\sqrt{\pi/2}-y^3/3$,  for a target of size $a$ which is much smaller than the typical distance $n$ between the reactants, Eq.~(\ref{ExpressionTMarkovianCentre}) reads
\begin{align}
	T=\int_0^{\infty}dt\left\{\frac{1}{[1-\phi^2]^{3/2}}-1\right\}.\label{WFMFCT}
\end{align} 
Changing of variables in (\ref{WFMFCT}) according to Eq.~(\ref{large_phi}), we obtain 
\begin{align}
	T=\tau_N \int_0^{\infty}d\tau\left\{\frac{1}{[1-\Phi^2]^{3/2}}-1\right\}\sim\tau_N\sim\tau_0n^{1/\gamma}
\end{align}
Note that the above integral converges because $\Phi(\tau)$ decays exponentially at large times, and behaves as $1-\tau^{\gamma}$ for small times, with $\gamma<2/3$. 
 
Now, the expression for MFCT in non-Markovian case has a similar structure as  Eq.~(\ref{ExpressionTMarkovianCentre}). The difference being in the first term, where $a^2\phi(t)^2$ in the exponent is replaced through a square of the reactive conformation $R_{\pi}(t)=\sum_{\lambda} b_{\lambda} m_{\lambda} e^{-\lambda t/\tau_0}$. Given that the analytic structure of the $m_{\lambda}$ involved is unknown ($m_{\lambda}$ are obtained from a set of self-consistent Eq. (\ref{EquationFirstMomentDim3AveragedSimplified2})), we cannot copy steps of WF and hence numerical evaluations are needed. For this the iterative procedures described below are of much help.

\section{Calculation of the coefficients $b_{\lambda}$ by decimation of Vicsek fractals} 

\begin{figure}
\begin{center}
\includegraphics[width=7cm]{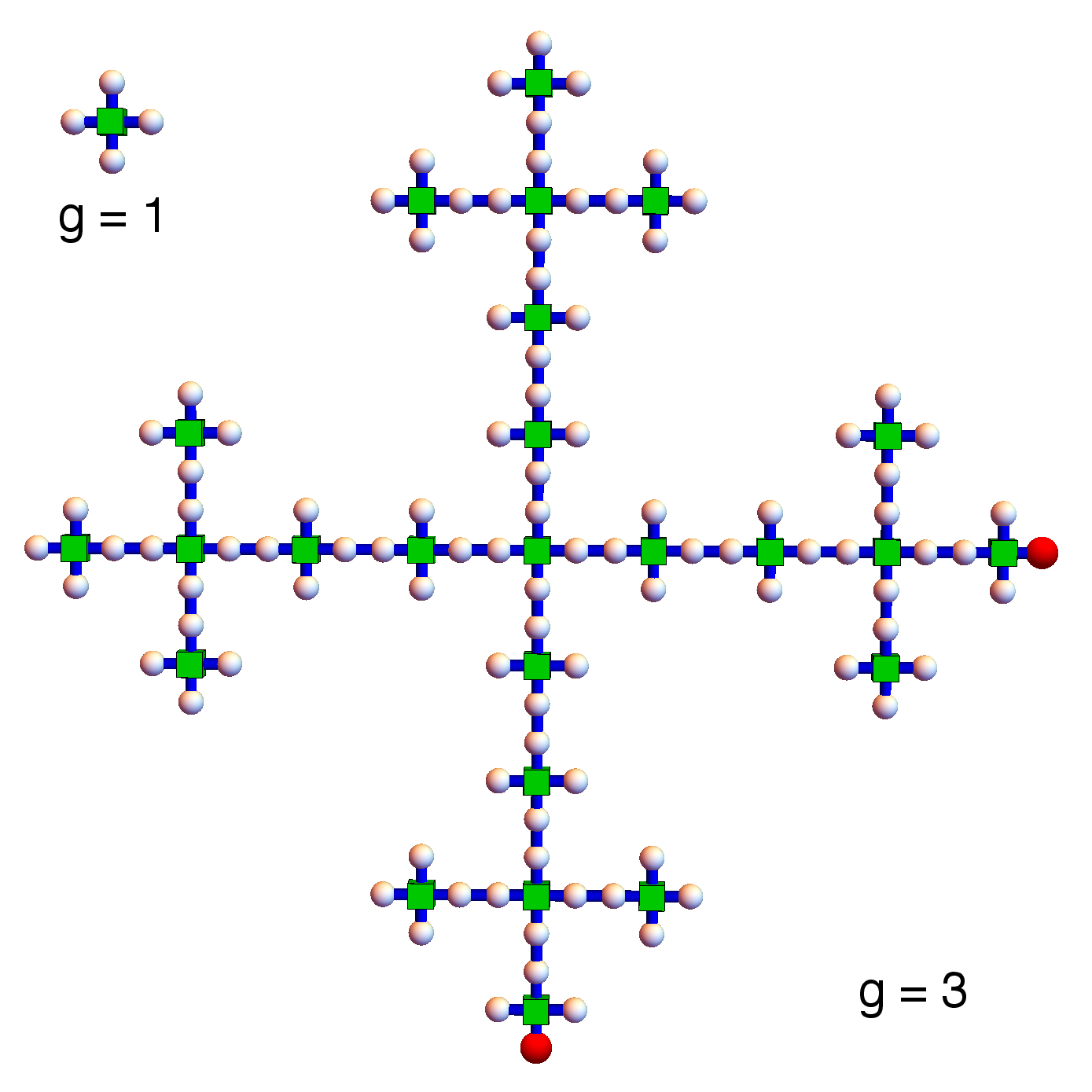}
\end{center}
\caption[kurzform]{Vicsek fractal (VF) of functionality $f=4$ and generation $g=3$. The reactive beads are colored in red. The picture shows only the topology of the structure, particular VF conformations may come up in vastly different geometric forms.}\label{VF}
\end{figure}

In this subsection we show for Vicsek fractals how to construct the matrices $\mathbf{W}_{\lambda}$ involved in Eq.~(\ref{b_lambda_final}). We adapt a decimation procedure that was already used in Ref. \cite{Blumen2004} for the computation of the eigenvalues of the dynamical matrix.  The topology of a Vicsek fractal is shown on Fig. \ref{VF}, where we have used two symbols for the beads: circles and squares. It is clear that removing all the circles from the structure, one obtains another Vicsek fractal but of a lower generation, formed by squares in Fig. \ref{VF}. Equivalently, one can say that the structure at the current generation is formed by the beads that were already part of the previous generation (the squares in Fig. \ref{VF}, their indices are denoted by greek letters, say $\mu$) and by "new" beads (the circles on Fig. \ref{VF}, their indices are denoted by latin letters, say by $k$). 

Now, consider an eigenvector $\ve[\Phi]^{(g)}=(\phi_1,...,\phi_N)$ of the dynamical matrix of a Vicsek fractal of generation $g$ which is associated to the eigenvalue $\lambda^{(g)}$. It is shown in Ref. \cite{Blumen2004} that, if $\lambda^{(g)}\ne\{0,1,f+1\}$, then necessarily the vector $\ve[\Phi]^{(g-1)}=\{\phi_{\mu}\}$ is itself an eigenvector of a Vicsek fractal of the former generation ($g-1$), where $\mu$ represents the ensemble of indices of the beads present at the former generation. The eigenvalue $\lambda^{(g-1)}$ to which $\{\phi_{\mu}\}$ is associated is then 
\begin{align}
	\lambda^{(g-1)}=\lambda^{(g)}(\lambda^{(g)}-3)(\lambda^{(g)}-f-1).  \label{CubicEquation}
\end{align}
If the values of $\{\phi_{\mu}\}$ are known for the sites present at generation $g-1$, the values of $\phi_k$ for the "new" sites, are deduced from the equations $\sum_{i=1}^N(A_{ki}-\lambda^{(g)}\delta_{ki})\Phi_i^{(g)}=0$ for all the new sites $k$. Hence, if one knows the matrix $W_{\lambda^{(g-1)}}^{(g-1)}$, then the matrix $W_{\lambda^{(g)}}$ at next generation is
\begin{align}
	\label{iterativeW}
\mathbf{W}^{(g)}_{\lambda^{(g)}}=
\begin{pmatrix}
\mathbf{W}^{(g-1)}_{\lambda^{(g-1)}} & \mathbf{0} \\
\text{Lines of } & A^{(g)}-\lambda^{(g)} I \ \text{ for the "new" sites $k$}
\end{pmatrix}.
\end{align}
This relation allows the iterative construction of the matrices $\ve[W]_{\lambda}$. Actually, since (\ref{CubicEquation}) is a cubic equation, an eigenvalue at generation $g-1$ will generate three eigenvalues at generation $g$ and the  3 associated matrices  $\ve[W]_{\lambda}$. 

The procedure needs to be initialized by expliciting the matrix $\ve[W]_1^{(g)}$ associated to the eigenvalue $\lambda=1$ at generation $g$. An eigenvector $\ve[\Phi]=(\phi_1,...,\phi_N)$ is associated to the eigenvalue $\lambda=1$ if and only if $(\mathbf{A}-\mathbf{I})\bf{\Phi}=0$. This leads to the equations for all the beads $\mu$ of functionality $f$:
\begin{align}
\phi_{\mu}=0,\label{eig1_f}
\hspace{0.4cm}\sum_{i\in \text{ NN of } \mu}\phi_{i}=0,
\end{align}
where NN means nearest-neighbors. Also, for each pair of sites of functionality 2, we obtain the $(f+1)^{g-1}-1$ equations
\begin{equation}\label{eig1_2}
	\phi_{i}-\phi_{j}=0.
\end{equation}
The equations 
(\ref{eig1_f},\ref{eig1_2}) are linearly independent and determine the eigensubspace related to $\lambda=1$. We can rewrite these equations under the matrix form $\ve[W]_1^{(g)}\ve[\Phi]=\ve[0]$, leading to the ready identification of the matrix $\ve[W]_1^{(g)}$. 

We note, that we did not consider the eigenvalues that are generated iteratively from the eigenvalue $\lambda=f+1$, which are non-degenerate. These eigenvalues are associated \citep{Blumen2004,Fuerstenberg2013} to eigenvectors $\{\Phi\}$ that all satisfy  $\phi_{q_1}=\phi_{q_2}$, where $q_1$ and $q_2$ are the indices of the reactive beads (the red beads on Fig.~\ref{VF}), located at the extremities of the structure in this work. Hence, based on Eqs. (\ref{Def_Vector_h}, \ref{DefBLambda}) we see that the coefficients $b_{\lambda}^2$ vanish for these non-degenerate eigenvalues, which do not need to be considered anymore. 
We also note that the eigenvalue $\lambda=0$, associated to the motion of center-of-mass, has no relevance in our problem concerned with intramolecular reactions.

\begin{figure}
\begin{center}
\includegraphics[width=7cm]{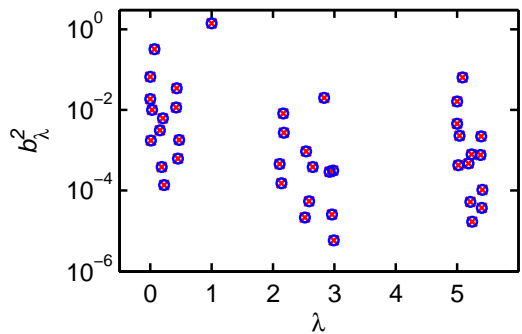}
\end{center}
\caption[kurzform]{Coefficients $b_{\lambda}^2$ for a Vicsek fractal of generation $g=4$ and functionality $f=4$. The reactive beads are at the extremities of the structure, as depicted on Fig.~\ref{VF}. The red crosses represent the results of a brute-force calculation (Eq.~(\ref{DefBLambda})) while the blue circles are obtained with a decimation procedure (Eq.~(\ref{b_lambda_final})).}\label{VF_check}
\end{figure}

Finally, the knowledge of all the matrices $\mathbf{W}_{\lambda}$ (which are very sparse) allows one to compute the coefficients $b_{\lambda}^2$ based on Eq.~(\ref{b_lambda_final}). We checked our iterative procedure by comparing our results to the direct computation issued from brute force diagonalization and the use of Eq. (\ref{DefBLambda}) (see Fig. \ref{VF_check})

\section{Calculation of the coefficients $b_{\lambda}$ for the dual Sierpi\'{n}ski gaskets by a decimation procedure. } 

\begin{figure}[h!]
 \includegraphics[width=7.5cm,clip]{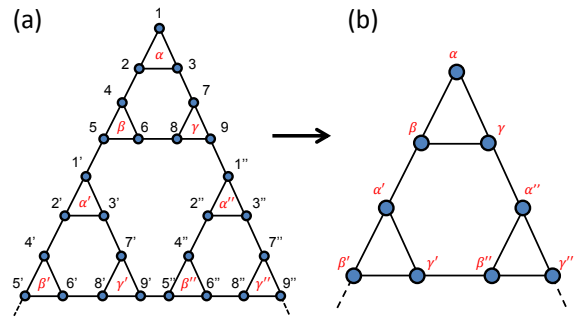}   
 \caption{(a) Topology of one part of a dual Sierpinski gasket. (b) decimated dual Sierpinski gasket. }
\label{PlotSierpinski}
\end{figure}

We now describe the decimation procedure for the dual Sierpinski gasket. We represented a part of such fractals on Fig. \ref{PlotSierpinski}a. As seen on this figure, the structure contains small triangles, and if one replaces each triangle by a new site, one obtains another dual Sierpinski gasket, at former generation (Fig. \ref{PlotSierpinski}b). This remark is at the basis of the decimation procedure.

Let us consider an eigenvector $\ve[\Phi]=(\phi_1,\phi_2,...)$ of the dynamical matrix $\ve[A]$ associated to the eigenvalue $\lambda$, which therefore satisfies 
\begin{align}
	\sum_{j=1}^N A_{ij}\phi_j=\lambda\phi_i. \label{84725}
\end{align} 
Let us consider another vector $\ve[\Psi]=(\psi_{\alpha},\psi_{\beta},...)$ (whose indices are the sites in the decimated structure), defined such as the coordinate $\psi_{\alpha'}$ is the sum of the values of $\phi_{i'}$ for the sites $i'$ in the triangle around $\alpha'$, and of the sites that are neighbors to these sites. For example, for the site $\alpha'$, we write (see the labeling on Fig.\ref{PlotSierpinski}a)
\begin{align}
&\psi_{\alpha'}=\phi_{1'}+\phi_{2'}+\phi_{3'}+\phi_{4'}+\phi_{7'}+\phi_5\label{DefPsiAlphaPrime},
\end{align}
For the sites at the boundary of the structure (\textit{i.e.} of functionality $2$), we modify the definition of $\psi$ by setting
\begin{align}
\psi_{\alpha}=2\phi_1+\phi_2+\phi_3+\phi_4+\phi_7.\label{DefPsiAlpha}
\end{align}
The relations (\ref{84725}) (written for the sites $1',2',...,9'$) and 
(\ref{DefPsiAlpha},\ref{DefPsiAlphaPrime}) (written for $\alpha,\beta,\gamma$) can be inverted to derive 
\begin{align}
	\phi_{i'}=\sum_{\mu=\{\alpha,\beta,\gamma\}} C_{i,\mu}\psi_{\mu'}, \label{95934}
\end{align}
where $C$ is a $9\times3$ matrix, of elements
\begin{align}
	C_{1\alpha}&=C_{5\beta}=C_{9\gamma}=L[-7-\lambda(\lambda-6)],\label{9583}\\
	C_{1\beta}&=C_{1\gamma}=C_{5\alpha}=C_{5\gamma}=C_{9\alpha}=C_{9\beta}=L,\\
	C_{2\alpha}&=C_{3\alpha}=C_{4\beta}=C_{6\beta}=C_{7\gamma}=C_{8\gamma}\nonumber\\
	&=L(-4+\lambda),\\
	C_{7\alpha}&=C_{3\gamma}=C_{6\gamma}=C_{4\alpha}=C_{2\beta}=C_{8\beta}=-L.\label{54215}
\end{align}
All other elements of $C$ are zero, and here $L$ is the constant  
\begin{align}
	L=[(\lambda-5)(\lambda-3)(\lambda-2)]^{-1}.\label{DefL}
\end{align}
Similarly, considering the 9 beads (1,2,...9) near the upper boundary of the structure, we obtain 
\begin{align}
	\phi_{i}=\sum_{\mu=\{\alpha,\beta,\gamma\}} C_{i,\mu}\psi_{\mu},
\end{align}
where the matrix $C$ is the same as in (\ref{95934}). Thus, we have obtained an inverse relation for the coordinates $\phi_i$ as a function of the vector defined on the decimated structure $\psi_{\alpha}$, which holds as soon as $\lambda\ne\{2,3,5\}$ (otherwise $L$ is infinite in (\ref{DefL})).
Furthermore, we obtain the relations
\begin{align}
[3-(5-\lambda)\lambda]\psi_{\alpha'}=\psi_{\beta'}+\psi_{\gamma'}+\psi_{\beta}\label{9585}
\end{align}
and similar relations hold for $\psi_{\beta},\psi_{\gamma}$, while for the site at the boundary, 
\begin{align}
	[2-(5-\lambda)\lambda]\psi_{\alpha}=\psi_{\beta}+\psi_{\gamma}. \label{85735}
\end{align}
Equations (\ref{9585},\ref{85735}) mean that $\ve[\Psi]$ is an eigenvector of the decimated structure, with the eigenvalue 
\begin{align}
	\tilde{\lambda}=(5-\lambda)\lambda.  \label{RecursiveRelationSG}
\end{align}
Equation (\ref{RecursiveRelationSG}) is an iterative relation between the eigenvalues at one generation ($\tilde{\lambda}$) and the eigenvalues at larger generation ($\lambda$), and is already known  \cite{Rammal,Cosenza}. 
Thus, an eigenvector $\ve[\Phi]$ associated to an eigenvalue $\lambda\ne\{2,3,5\}$ can be expressed as a product $\ve[C]_{\lambda}\ve[\Psi]$, where $\ve[C]_{\lambda}$ is the matrix defined in   (\ref{9583}-\ref{54215}), and $\ve[\Psi]$ is an eigenvector of the decimated structure (at generation $g-1$) associated to the eigenvalue $\lambda^{(g-1)}=(5-\lambda)\lambda$. Let us iterate the procedure and consider $s$ decimation steps, until one reaches an eigenvalue $\lambda^{(g-s)}=\{3,5\}$: then an eigenvector $\ve[\Phi]$ is expressed as $\ve[\Phi]=\ve[B]\ve[\Psi]^{(g-s)}$, where $\ve[\Psi]^{(g-s)}$ is an eigenvector at generation $g-s$, and $\ve[B]=\Pi_{s'=0}^{s-1}\ve[C]_{\lambda^{(g-s')}}$. 

Let $\mathbf{W}_{3}$ be the matrix of vectors which span the subspace  orthogonal to all the eigenvectors related to the eigenvalue $\lambda=3$ at generation $g-s$. Then, in the basis related to the eigenvalue $\lambda=3$ at generation $g-s$ Eq.(\ref{b_lambda_final}) transforms to
\begin{align}\label{b_lambda_dsg}
b_{\lambda}^2=&(\ve[h]|\mathbf{B}(\mathbf{B}^T\mathbf{B})^{-1}\mathbf{B}^T|\ve[h])\nonumber\\
&-(\ve[h]|\mathbf{B}\mathbf{W}_{3}^T(\mathbf{W}_{3}\mathbf{B}^T\mathbf{B}\mathbf{W}_{3}^T)^{-1}\mathbf{W}_{3}\mathbf{B}^T|\ve[h]).
\end{align}

\begin{figure}
\begin{center}
\includegraphics[width=7.5cm]{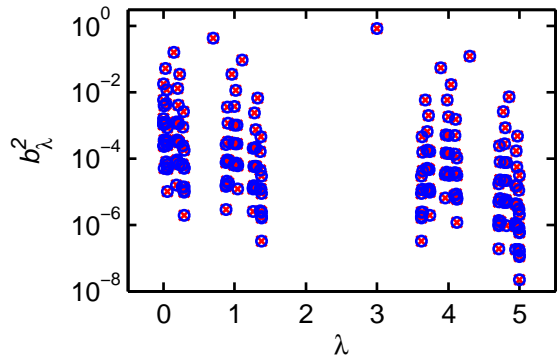}
\end{center}
\caption[kurzform]{Coefficients $b_{\lambda}^2$ for a dual Sierpi\'{n}ski gasket of generation $g=8$. The reactive beads are at extremities, as depicted on Fig.1 in the main text. Red crosses represent the results of the brute-force calculation (Eq.~(\ref{DefBLambda})) and blue circles of the decimation procedure  (Eq.~(\ref{b_lambda_dsg})).}\label{DSG_check}
\end{figure}

The eigenvalue $\lambda=3$ plays a fundamental role in the whole decimation procedure. The eigenvectors related to this eigenvalue are the so-called symmetric modes~\cite{Cosenza}. This means that for each eigenvector  $\mathbf{\Phi}$ associated to $\lambda=3$, the following relations hold~\cite{Cosenza}:
\begin{equation}\label{eig3_eq1}
\phi_i+\phi_j+\phi_k=0,
\end{equation}
for each triplet  $(i,j,k)$ of nearest-neighboring beads forming the smallest triangles of the structure (see Fig. \ref{PlotSierpinski}). Furthermore, we have also \cite{Cosenza} 
\begin{equation}\label{eig3_eq2}
\phi_n=\phi_m,
\end{equation}
for each pair $(n,m)$ of nearest-neighboring beads connecting the smallest triangles. Equations (\ref{eig3_eq1},\ref{eig3_eq2}) are linearly independent and determine the eigensubspace of $\ve[A]$ associated to $3$, they can therefore be written in matrix form $\ve[W]_3\ve[\Phi]=\ve[0]$, from which the (sparse) matrix $\ve[W]_3$ reads.

On the contrary, the eigenvalue $\lambda=5$, associated to antisymmetric modes \cite{Cosenza}, does not play any role in our problem: when the reactive beads are located at the extremities of the structure (Fig.~1b in the main text), the coefficients $b_{\lambda}^2$ vanish for symmetry reasons for $\lambda=5$ and for any eigenvalue generated from the eigenvalue $5$. 

We implemented a recursive algorithm which calculates the coefficients $b_{\lambda}$ with our decimation procedure. The results were checked by comparing them to the output of the brute force diagonalization of the dynamical matrix, one sees on Fig. \ref{DSG_check} that our procedure gives the correct results. We used our decimation approach to treat much larger structures and to generate Fig. 3 of the main text.

\bibliography{bibfile}

\bibliographystyle{apsrev4-1}

\end{document}